# Structural prediction of Fe-Mg-O compounds at Super-Earth's pressures

Yimei Fang[1], Yang Sun[1,2*], Renhai Wang[3], Feng Zheng[4], Feng Zhang[2], Shunqing Wu[1*], Cai-Zhuang Wang[2], Renata M. Wentzcovitch[5,6,7,8*], Kai-Ming Ho[2]

[1]Department of Physics, OSED,
Key Laboratory of Low Dimensional Condensed Matter Physics
(Department of Education of Fujian Province)
Jiujiang Research Institute, Xiamen University, Xiamen 361005, China.
[2]Department of Physics, Iowa State University, Ames, Iowa 50011, USA
[3]School of Physics and Optoelectronic Engineering, Guangdong University of Technology, Guangzhou 510006, People's Republic of China
[4]School of Science, Jimei University, Xiamen 361021, China
[5]Department of Applied Physics and Applied Mathematics, Columbia University, New York, NY, 10027, USA
[6]Department of Earth and Environmental Sciences, Columbia University, New York, NY, 10027, USA
[7]Data Science Institute, Columbia University, New York, NY, 10027, USA
[8]Lamont–Doherty Earth Observatory, Columbia University, Palisades, NY, 10964, USA

*Corresponding authors: yangsun@xmu.edu.cn (Y. S.); wsq@xmu.edu.cn (S.Q.Wu); rmw2150@columbia.edu (R.M.W.)

**Abstract**

Terrestrial exoplanets are of great interest for being simultaneously similar to and different from Earth. Their compositions are likely comparable to those of solar-terrestrial objects, but their internal pressures and temperatures can vary significantly with their masses/sizes. The most abundant non-volatile elements are O, Mg, Si, Fe, Al, and Ca, and there has been much recent progress in understanding the nature of magnesium silicates up to and beyond ~3 TPa. However, a critical element, Fe, has yet to be systematically included in materials discovery studies of potential terrestrial planet-forming phases at ultra-high pressures. Here, using the adaptive genetic algorithm (AGA) crystal structure prediction method, we predict several unreported stable crystalline phases in the binary Fe-Mg and ternary Fe-Mg-O systems up to pressures of 3 TPa. The analysis of the local packing motifs of the low-enthalpy Fe-Mg-O phases reveals that the Fe-Mg-O system favors a *BCC* motif under ultra-high pressures regardless of chemical composition. Besides, oxygen enrichment is conducive to lowering the enthalpies of the Fe-Mg-O phases. Our results extend the current knowledge of structural information of the Fe-Mg-O system to exoplanet pressures.

## Introduction

The past two decades have witnessed a rapid increase in the number of discovered exoplanets, among which terrestrial-type exoplanets with masses under $10M_{\oplus}$ ($M_{\oplus}$: Earth's mass), frequently referred to as "super-Earth," are of particular interest for their similarities to and differences from Earth and their potential to host life. Among other factors, this life-host potential depends on its internal structure, particularly the existence of a planetary core able to produce a radiation-shielding magnetic field at the surface. The primary non-volatile elements are expected to be the same as those of solar terrestrial planets, i.e., O, Mg, Si, Fe, Al, and Ca [1]. Their internal structure is also expected to consist of an iron-rich metallic core and a surrounding silicate mantle (see, e.g., [2, 3]), just like the solar terrestrial ones. However, the internal pressure in a large terrestrial planet can be much higher, e.g., approximately 1.2 TPa at the core-mantle boundary (CMB) of a planet with $\sim 12M_{\oplus}$ ($M_{\oplus}$: Earth's mass) [2].

Currently, considerable theoretical efforts have focused on investigating the post-post-perovskite (post-PPV) phase transitions in the Mg-Si-O system [2, 4-8] at exoplanet conditions. However, Fe is one of the most abundant elements on terrestrial planets, and it must combine with the other rock-forming elements to form Fe-bearing compounds. Besides, given that high pressure facilitates the mixture of Fe/Mg compounds [9-12], it is reasonable that abundant Fe-Mg-O ternary phases should exist under exoplanetary conditions. Although the thermodynamic and spin crossover properties [13-15] in ferropericlase (i.e., B1-type $Mg_{1-x}Fe_xO$) at Earth's lower-mantle conditions and super-Earth'smantle conditions (B2-type $Mg_{1-x}Fe_xO$) [16-18] have been extensively studied, relatively less attention has been paid to the Fe-Mg-O phases with other metal-to-oxygen ratios at exoplanetary pressures.

Here, using the Adaptive Genetic Algorithm (AGA) [19] we perform structure predictions for the ternary Fe-Mg-O phases with variable compositions at 1 TPa and 3 TPa. Two unreported compounds, i.e., $FeMgO_3$ and $FeMg_3O_4$, are found to be stable. The analysis of the local packing motifs of the Fe-Mg-O phases with lower enthalpies leads us to conclude that the Fe-Mg-O system favors a *BCC-type* motif under ultra-high pressures regardless of chemical composition. Additionally, we find that an O-rich environment is beneficial to lowering the enthalpies of the Fe-Mg-O phases. While the effects of temperature are not included in the current work, our present study provides ample structural and motif information for future analysis of the thermodynamic properties of these compounds at high-pressure and high-temperature conditions.

**Computational Methods**

The AGA method [19] combines the efficiency of structure exploration by classical potential calculations and the accuracy of density functional theory (DFT) calculations adaptively and iteratively. In the current AGA searches for Fe-Mg-O phases, unit cells containing up to 4 f.u. were used, and initial atomic positions were randomly generated without any assumption on the symmetry. Structure searches were performed at 1TPa and 3TPa, and the enthalpy was used as the selection criteria for optimizing the candidate structures pool. We kept the pool size in the current searches to 128 structures. At each GA generation, 32 (*i.e.,* a quarter of the pool size) new structures are produced from the parent structure pool through the mating procedure described in [20], and are updated to the candidate pool as parent structures for the next generation. The structure search was carried out for 600 consecutive GA generations using a specific auxiliary interatomic potential. After the GA search, 16 lowest-enthalpy structures were selected for first-principles calculations to obtain the enthalpies, forces, and stresses to re-adjust the potential parameters of the classical auxiliary potential for the next GA search. We performed 60 adaptive iterations to obtain the final structures for a given chemical composition.

Here, interatomic potentials based on the embedded-atom method (EAM) [21] were chosen as the classical auxiliary potential. Within the EAM, the total energy of an *N*-atom system takes the form of

$$E_{total} = \frac{1}{2}\sum_{i,j(i\neq j)}^{N} \phi(r_{ij}) + \sum_i F_i(n_i), \tag{1}$$

where $\phi(r_{ij})$ is the pair repulsion between atoms $i$ and $j$ with a distance of $r_{ij}$. $F_i(n_i)$ is the embedded term with electron density term $n_i = \sum_{j\neq i}\rho_j(r_{ij})$ at the site occupied by atom $i$. The fitting parameters in the EAM formula for the Fe-Mg-O system are determined as follows: the parameters for Fe-Fe and Mg-Mg interactions were taken from the literature [22], and the other pair interactions (*i.e.,* O-O, Fe-Mg, Fe-O, and Mg-O) were modeled by the Morse function,

$$\phi(r_{ij}) = D\left[e^{-2\alpha(r_{ij}-r_0)} - 2e^{-\alpha(r_{ij}-r_0)}\right], \quad (2)$$

where $D, \alpha, r_0$ are the fitting parameters. For O atoms, the density function was modeled by an exponentially decaying function

$$\rho(r_{ij}) = \alpha\, exp[-\beta(r_{ij} - r_0)], \quad (3)$$

where $\alpha$ and $\beta$ are the fitting parameters; the embedding function was expressed by

$$F(n) = F_0[1 - \gamma\, In\, n]n^{\gamma}. \quad (4)$$

with fitting parameters $F_0, \gamma$, as proposed by Benerjea and Smith in ref. [23]. For Fe and Mg, the parameters of the density function and embedding function were also taken from ref. [22].

During the AGA run, the potential parameters were adjusted adaptively by fitting to the selected structures' DFT-calculated enthalpies, forces, and stresses. The fitting procedure was conducted by adopting the force-matching method with a stochastic simulated annealing algorithm as implemented in the POTFIT code [24, 25].

The first-principles calculations were carried out within density functional theory (DFT) as implemented in the Quantum ESPRESSO (QE) code [26, 27]. The exchange-correlation functional was treated with the non-spin-polarized generalized-gradient approximation (GGA) as parameterized by the Perdew-Burke-Ernzerhof formula (PBE)[28]. A kinetic-energy cutoff of 50 Ry for wave functions and 500 Ry for potentials were used. The Brillouin-zone integration was performed over a k-point grid of $2\pi\times0.03$ $Å^{-1}$ in the structure refinement. Convergence thresholds are 0.01 eV/Å for the atomic force, 0.05 GPa for the pressure, and $1\times10^{-5}$ eV for the total energy.

## Results and Discussion

### Phase stability

We systematically investigate the phase stabilities of several stoichiometric $Fe_xMg_yO_z$ compounds predicted by the AGA search. We assign integer values from 1 to 4 to x, y, and z. This results in 64 different Fe, Mg, and O compositions. After removing redundant compositions, we searched 55 distinct stoichiometries of $Fe_xMg_yO_z$ with up to 56 atoms at 1 TPa and 3 TPa. The formation enthalpy per atom ($H_f$) of a $Fe_xMg_yO_z$ phase at a given pressure is expressed as:

$$H_f = \frac{H_{Fe_xMg_yO_z} - (xH_{Fe} + yH_{Mg} + zH_O)}{x + y + z}, \quad (5)$$

where $H_{Fe_xMg_yO_z}$ stands for the enthalpy of the $Fe_xMg_yO_z$ phase, $H_{Fe}$, $H_{Mg}$, and $H_O$ for the enthalpies of pure Fe, Mg, and O elements, respectively. We can construct the ternary convex hull based on the formation enthalpies of all the searched phases. Here, the convex hull is a hypersurface in compositional space that connects all elementary, binary, and ternary compounds that are thermodynamically stable against decomposition [29]. A $Fe_xMg_yO_z$ phase is identified as stable if it lies on the vertex of the convex hull. If a $Fe_xMg_yO_z$ phase is above the convex hull, it is metastable and should decompose to the stable phases under thermodynamical equilibrium.

To construct the ternary convex hull, we first address the ground-state structures of elementary and binary phases on the phase diagram. Previous studies have investigated elementary and binary phases consisting of Fe, Mg, and O. It has been reported that at 500 GPa, elementary Fe adopts the *hcp* structure [30]. Our calculations show that the *hcp* phase of Fe remains as the ground state at both 1 TPa and 3 TPa compared to the *bcc* phase. In the pressure range of our interest, Mg undergoes a phase transition from the simple hexagonal structure (*sh*) to the simple cubic (*sc*) structure at 1.07 TPa [8, 31, 32]. Consequently, the *sh* and *sc* structures are the reference phases at 1 TPa and 3 TPa, respectively. Elementary oxygen exhibits a variety of phases within a range of pressures from 0.25 TPa to 10 TPa [33]. By performing structural optimization for these phases at 1 TPa and 3 TPa, we find that the $I4_1/acd$ phase is the ground-state phase at 1 TPa and the $Cmcm$ phase with zigzag-chain-like structure is the ground state at 3 TPa. Note that Ref [33] indicates $I4_1/acd$ is a metastable state at 1 TPa and it only becomes the ground state at 2 TPa. This discrepancy may stem from the different pseudopotentials used in the calculations. Our pseudopotential was generated by the Vanderbilt′s method [34]. It had been tested in a few studies at high pressures [4, 7], while Ref [33] used the projector augmented wave (PAW) pseudopotentials shipped with VASP software.

For the reference binary phases, the stable Mg-O compounds under high pressures have been extensively explored by Niu and Zhu [8, 31]. MgO, $MgO_2$, and $MgO_3$ phases are found to be stable ground states at 1 TPa. While the MgO and $MgO_3$ structures remain stable at 3 TPa, the $MgO_2$ phase decomposes into MgO and $MgO_3$ at 1.43 TPa. Therefore, we use MgO, $MgO_2$, and $MgO_3$ as the reference at 1 TPa, while only MgO and $MgO_3$ at 3TPa. Weerasinghe *et al.* [30] predicted several stable structures with different stoichiometries for up to 500 GPa for the Fe-O system. By extending their calculation to 1 TPa and 3 TPa, we found four of their predicted structures, i.e., FeO, $FeO_2$, $FeO_4$, and $Fe_2O$ are still stable ground states at 1TPa, while only FeO, $Fe_2O$, and $FeO_2$ remain ground states up to 3 TPa [35]. In addition, we recently identified two different stable $Fe_3O_5$ phases at 1 TPa and 3TPa [35]. Gao *et al.* [10] have recently predicted six unexpected phases at 360 GPa for the Fe-Mg system. Their $Fe_2Mg$, $FeMg_2$, and $FeMg_3$ phases remain stable at 1 TPa, but all become unstable at 3 TPa [12]. Ref. [12] gives a detailed discussion of the stabilities of the binary Fe-Mg phases. Here we use the structures of these unreported stable Fe-Mg compounds, i.e., $Fe_2Mg_3$, FeMg, $Fe_3Mg_2$, and $FeMg_2$, as the binary references under 3 TPa. Details of the space groups and lattice parameters of the above-mentioned elementary and binary phases are listed in Supplementary Table S1 for 1 TPa and Table S2 for 3TPa, respectively. The crustal structures of all the stable elementary and binary phases on the convex hull at 1 TPa and 3 TPa are displayed in Fig. S1 and Fig. S2, respectively.

Based on these elementary and binary references, we construct the Fe-Mg-O system's ternary phase diagrams and convex hull. As depicted in Fig. 1(a), the ternary phase diagram exhibits two stable stoichiometric Fe-Mg-O compounds at 1TPa, namely, $FeMgO_3$ and $FeMg_3O_4$. These two structures remain the ground state at 3 TPa even though the binary references differ, as shown in Fig. 1(b).

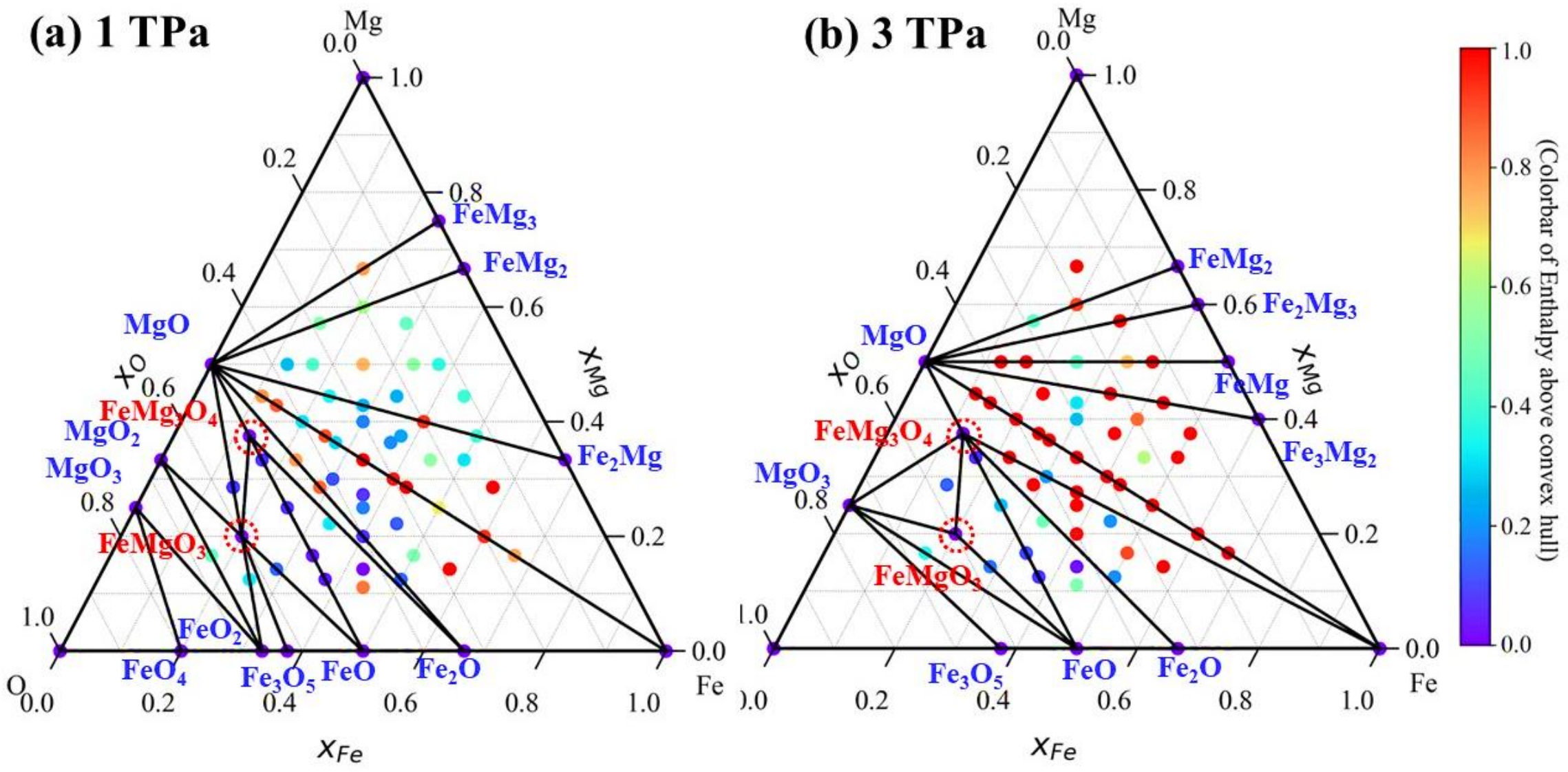

Fig. 1. Ternary phase diagram of the Fe-Mg-O system at (a) 1 TPa and (b) 3 TPa. The points marked by red dashed circles represent the ground-state $Fe_xMg_yO_z$ phases. The colors of solid circles indicate the relative formation enthalpy in eV/atom to the convex hull.

**Structural geometry**

Figure 2 shows the predicted ternary $Fe_xMg_yO_z$ phases' ground-state structures (see Supplementary Tables S3 and Table S4 for crystallographic details). The oxygen-rich $FeMgO_3$ phase is an orthorhombic structure with $Pnma$ symmetry. As can be seen in Fig. 2(a), each Fe atom in the $FeMgO_3$ phase is coordinated to eight oxygen atoms, forming the Fe-centered $FeO_8$ cubes, while each Mg atom is coordinated to ten oxygen atoms, forming the $MgO_{10}$ polyhedrons. By sharing their faces, the $FeO_8$ and $MgO_{10}$ clusters constitute the $Pnma$ phase of $FeMgO_3$. The $FeMg_3O_4$ phase adopts a cubic structure with $Im\bar{3}m$ symmetry. In the $FeMg_3O_4$ lattice (2f.u. shown in Fig.2(b)), Mg atoms occupy the edge centers and face centers, while Fe atoms are located at the corners and body centers. Both Mg and Fe are in the environment of $FeO_8$/$MgO_8$ cubes. To check if there is any disordered Mg/Fe occupation, we generate several structures with the same oxygen occupation but randomly permutated metal sites with Fe/Mg. We optimize these structures and find these structures all have larger enthalpy than the ground-state structure from the AGA search. The enthalpy difference ranges from 66 to 325 meV/atom.

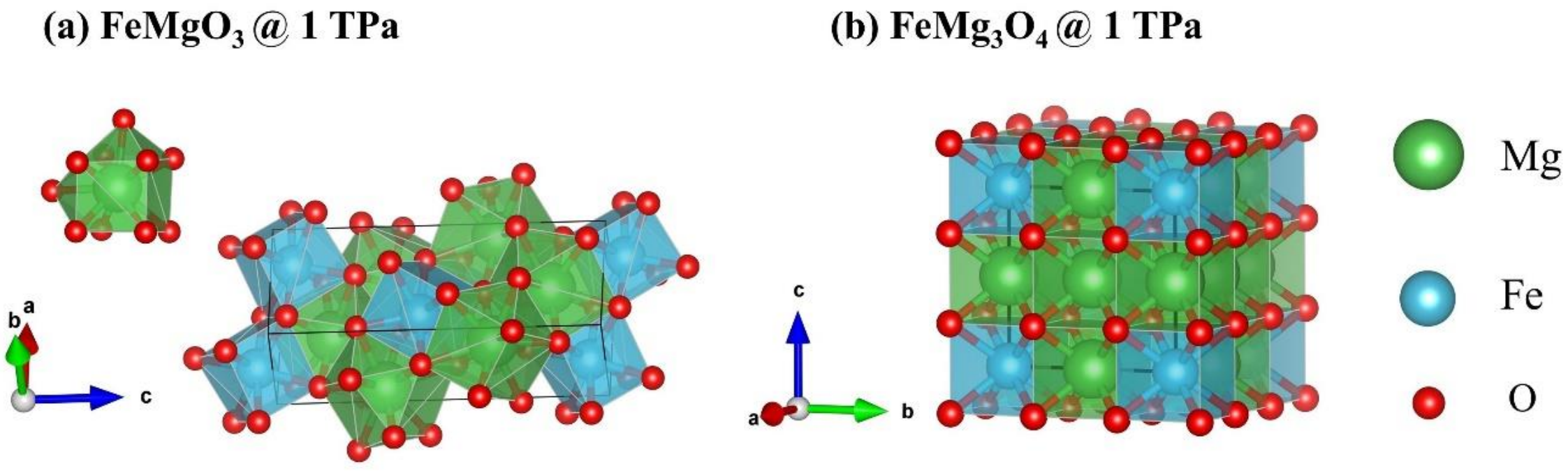

Fig. 2. Crystal structures of stable $Fe_xMg_yO_z$ phases: (a, b) $Pnma$ $FeMgO_3$ $and$ $Im\bar{3}m$ $FeMg_3O_4$.

**Local packing motifs**

Since the effect of temperature is not considered in this work, other structures metastable at T = 0K may become stable at high temperatures. For this reason, we also investigate the AGA-searched $Fe_xMg_yO_z$ phases with relative enthalpies ($\Delta H$) up to 0.8 eV/atom (~9000K) above the convex hull. These structures provide better statistics to explore further the impact of high pressure on the Fe-Mg-O system. We employ the cluster alignment (CA) method [36-38] to classify these metastable structures based on their local packing motifs. The clusters here are defined by a center atom and its first-shell neighbor atoms. In the CA analysis, a root mean squared deviation (RMSD) was used to describe the dissimilarity between an as-extracted cluster and the perfect template cluster, as

$$S = min_{\alpha} \sqrt{\frac{1}{n}\sum_{i=1}^{n}\frac{(\boldsymbol{r}_{ci} - \alpha\boldsymbol{r}_{ti})^2}{(\alpha\boldsymbol{r}_{ti})^2}}$$

where $n$ denotes the number of neighboring atoms in the template; $\boldsymbol{r}_{ci}$ and $\boldsymbol{r}_{ti}$ represent the atomic positions in the aligned clusters and template, respectively, and $\alpha$ is the optimal scaling parameter of the bond lengths in the template. Each as-extracted cluster from the sample is aligned to *FCC*, *BCC*, *OCT* (octahedron), *BCT* (body-centered tetragonal), and *HCP* templates, which are the most common motifs in the Fe-O and Mg-O system [8, 30, 31] as shown in Fig. S3. The motif with the minimum RMSD out of these five templates is assigned to the cluster. If the lowest RMSD of the five templates are still higher than a criterion of 0.125, the cluster is classified as an unrecognized type (noted as 'others' in the figures). This criterion is set for possible distortion of the perfect motifs in the crystal structures.

With the classified local motifs, Fig. 3 shows the scatter plot containing the information on the relative enthalpies with respect to the convex hull, the atomic volumes, the local packing motifs and the oxygen concentrations for the stable and metastable phases. A similar plot with the mass densities is presented in Supplementary Fig. S4 for the potential interest in the densities of these structures. Several conclusions can be drawn from Fig. 3. First, the averaged atomic volume directly correlates to the oxygen concentration. Higher oxygen concentration generally results in a smaller averaged atomic volume. Second, when comparing the low-energy structures among different oxygen concentrations, we find the oxygen-rich phases tend to be closer to the convex hull than the oxygen-poor phases at 1 TPa. A similar trend can be found at 3 TPa that the oxygen-poor phases have much higher energies above convex hull, as shown in Fig. 3(c-d). By comparing the Fe- and Mg-centered motifs in Fig. 3, we find that Fe and Mg have very similar local motifs in most phases, even though they can have different coordination numbers. For example, while the Mg atoms in $FeMgO_3$ in Fig. 2(a) have higher coordination than Fe, it still adopts a distorted *BCC* motif. Therefore, CA provides more accurate descriptions of the local packing motifs than the coordination number. Finally, the *BCC*-type clusters dominate lower-energy structures, especially in the oxygen-rich low-enthalpy phases at both 1TPa and 3TPa. This is consistent with the fact that both MgO and FeO adopt the B2 structure (BCC motif) under high pressures. The oxygen-poor phases with higher enthalpies show a mixture among *BCC*-, *BCT*-, *FCC*-, and *HCP*-type clusters. It should be noted that although the octahedron motif is the only local packing motif in ferropericlase ($Mg_{1-x}Fe_xO$) at lower pressures within 100 GPa, it is very rare under current pressure conditions.

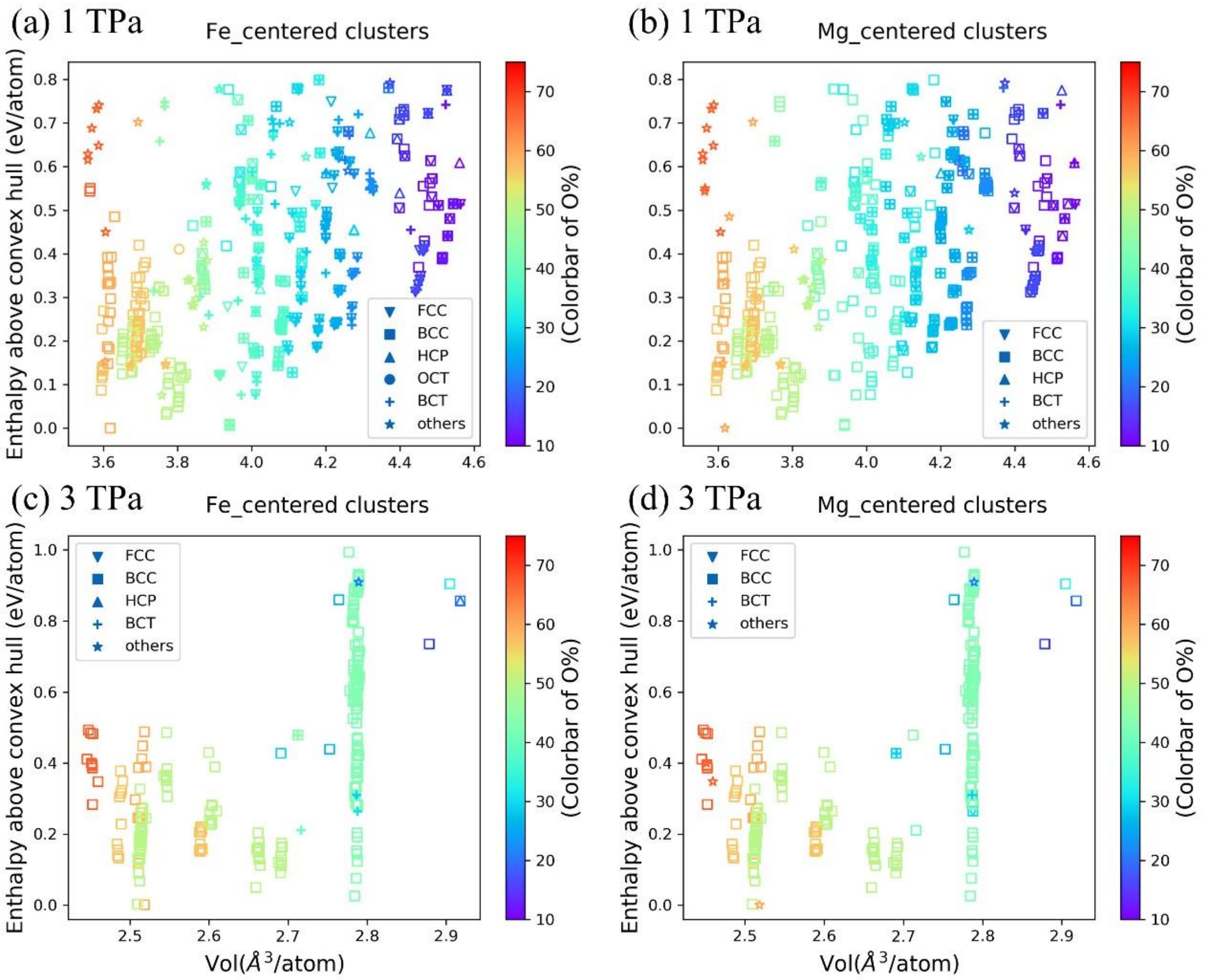

Fig. 3 The relative enthalpies of our predicted $Fe_xMg_yO_z$ structures as a function of their volumes, where the symbols indicate the local packing motifs, and the colors denote the oxygen concentration. (a, b) are for clusters with Fe and Mg as central atoms at 1 TPa, respectively, while (c, d) correspond to (a, b) at 3 TPa.

## Conclusions

In summary, we have investigated the structures and motifs of ternary $Fe_xMg_yO_z$ phases predicted by the adaptive genetic algorithm at 1 TPa and 3 TPa. Two stable phases, namely $FeMgO_3$ and $FeMg_3O_4$, are identified. The cluster alignment analysis reveals that BCC is the most favored packing motif for low-enthalpy Fe-Mg-O phases at these pressures. We also find that the oxygen-rich $Fe_xMg_yO_z$ phases generally have enthalpies closer to the convex hull than the oxygen-poor phases, suggesting that Mg-Fe cations are fully oxidized or nearly so at these high pressures and low temperatures. In this work, we provide a comprehensive structure database to assist future efforts to study the high-pressure structural behaviors of Fe-Mg-O compounds.

**Acknowledgments**

The Xiamen University's High-Performance Computing Center is acknowledged for its computational resources. Shaorong Fang and Tianfu Wu from Information and Network Center of Xiamen University are acknowledged for their help with GPU computing. Work at Iowa State University and Columbia University was supported by the National Science Foundation awards EAR-1918134 and EAR-1918126. Work at Guangdong University of Technology was supported by the Guangdong Basic and Applied Basic Research Foundation (Grant No. 2021A1515110328 & 2022A1515012174) and the Guangdong Natural Science Foundation of China (Grant No. 2019B1515120078).